\documentclass[showpacs,amsmath,amssymb,pra,superscriptaddress,floatfix,twocolumn]{revtex4-1}

\usepackage{longtable}
\usepackage{amssymb, graphicx,subfigure}
\usepackage{enumerate,tensor,xfrac}
\usepackage{enumerate}
\usepackage{nicefrac}
\usepackage{amsmath,bm}
\usepackage{hyperref}
\usepackage{epsfig}
 \usepackage{relsize}

\usepackage[per-mode=symbol]{siunitx}   
\usepackage[version=3]{mhchem}

\DeclareSIUnit\intensity{\watt\per\centi\meter\squared}
\DeclareSIUnit\fieldstrength{\volt\per\centi\meter}
\DeclareSIUnit\kfieldstrength{k\volt\per\centi\meter}
\DeclareSIUnit\magneticstrength{T}
\newcommand{\ie}{i.\,e.}%

\makeatletter
\def\subsubsection{\@startsection{subsubsection}{3}{10pt}{-1.25ex plus -1ex minus -.1ex}{0ex plus 0ex}{\normalsize\bf}}
\def\paragraph{\@startsection{paragraph}{4}{10pt}{-1.25ex plus -1ex minus -.1ex}{0ex plus 0ex}{\normalsize\textit}}
\renewcommand\@biblabel[1]{#1}
\renewcommand\@makefntext[1]%
{\noindent\makebox[0pt][r]{\@thefnmark\,}#1}
\DeclareRobustCommand\onlinecite{\@onlinecite}
\def\@onlinecite#1{\begingroup\let\@cite\NAT@citenum\citealp{#1}\endgroup}
\def\tagform@#1{\maketag@@@{\ignorespaces#1\unskip\@@italiccorr}}
\let\orgtheequation\theequation
\def\theequation{(\orgtheequation)}
\makeatother

\newcommand{\ry}{Rydberg }

\def\jpb#1#2#3{J.~Phys.~B~{\bf #1},\ #2\ (#3)}

\def\pra#1#2#3{Phys.~Rev.~A~{\bf #1},\ #2\ (#3)}

\def\prl#1#2#3{Phys.~Rev.~Lett.~{\bf #1},\ #2\ (#3)}

\begin{document}

\title[]{Ultralong-range triatomic Rydberg molecules in an electric field}

\author{Javier Aguilera Fern\'andez}
\affiliation{Instituto Carlos I de F\'{\i}sica Te\'orica y Computacional,
and Departamento de F\'{\i}sica At\'omica, Molecular y Nuclear,
  Universidad de Granada, 18071 Granada, Spain}
  
\author{Peter Schmelcher}

\affiliation{Zentrum f\"ur Optische Quantentechnologien, Universit\"at
  Hamburg, Germany} 
  \affiliation{The Hamburg Center for Ultrafast Imaging, Luruper Chaussee 149, 22761 Hamburg, Germany}

  \author{Rosario Gonz\'alez-F\'erez}
\email{rogonzal@ugr.es}
\affiliation{Instituto Carlos I de F\'{\i}sica Te\'orica y Computacional,
and Departamento de F\'{\i}sica At\'omica, Molecular y Nuclear,
  Universidad de Granada, 18071 Granada, Spain}

\vspace{10pt}
%\begin{indented}
%\item[]February 2014
%\end{indented}

\begin{abstract}
We investigate the electronic structure  of a triatomic Rydberg
molecule formed by a Rydberg atom and two neutral ground-state atoms. 
Taking into account the $s$-wave and $p$-wave interactions we perform electronic
structure calculations and analyze the adiabatic electronic potentials
evolving from the Rb$(n=35, l\ge 3)$ Rydberg degenerate manifold. 
We hereby focus on three different classes of geometries of the \ry molecules,
including symmetric, asymmetric and planar configurations.
The  metamorphosis of these potential energy surfaces in the presence of an external electric field
is explored.
 \end{abstract}

%s\submitto{\JPB}
\maketitle

\section{Introduction}
\label{sec:introduction}
In recent years, ultralong-range \ry molecules (ULRM) formed when a ground-state atom is bound to a \ry  
atom have developed into a field of intense 
research~\cite{bendkowsky09,bendkowsky10,gaj14,krupp14,gaj15,pfau15}.
These ultralong-range species, which were predicted theoretically back in 2000~\cite{greene00}, 
exhibit an exotic binding mechanism based on the low-energy collisions between a Rydberg electron and a ground-state atom.
The elastic scattering is typically described by $s$- and $p$-wave Fermi-type 
pseudopotentials~\cite{fermi34,Omont},
and leads to the unusual oscillatory behavior of the corresponding adiabatic potential energy curves.
The polar (trilobite) molecular states emerging from the near degenerate high angular momentum atomic Rydberg manifold 
possess a huge dipole moment whereas the non-polar molecular states emerging, e.g., from quantum defect split  
$s-$states exhibit only a minor dipole moment~\cite{Li11}. Their vibrational energies are in the GHz
and MHz regimes, respectively~\cite{greene00,hamilton02}. Apart from the Rb ULRMs within the above-mentioned
experiments, Cesium blue detuned ULRMs~\cite{Tallant12} and Strontium ULRMs~\cite{Killian15} from divalent atomic
systems have been prepared and explored. The detection of the singlet/triplet hyperfine structure and scattering
channels has been very recently performed in~\cite{Sassmannshausen15}. Due to the sensitivity of these \ry molecules to 
external fields, their electronic structure, molecular geometry and rovibrational dynamics could be
controlled and manipulated easily by using weak static magnetic and electric fields or laser 
fields~\cite{lesanovsky07,kurz12,kurz13,kurz14}. Very recently a selective excitation of rovibrational 
molecular states with a 
variable degree of alignment and antialignment has been demonstrated experimentally by using a magnetic 
field~\cite{krupp14}.

Adding more ground-state atoms to the \ry cloud, polyatomic ultralong-range molecules could be formed.
However, very little is known about the structure and properties of these species.
In Ref.~\cite{rost06} the emphasis was put on the splitting of the energy levels and the construction of symmetry-adapted orbitals.
While a repulsive interaction does not support bound states, it is shown in Ref.~\cite{rost09}
that adding a second ground-state atom, a long-range bound triatomic molecule becomes possible.
Recently Rydberg trimers and excited dimers bound by quantum reflection have been studied experimentally
and theoretically~\cite{bendkowsky10}. We note that experimentally the transition from a \ry dimer to a 
\ry polyatomic has been achieved by increasing the principal quantum number of the \ry state~\cite{gaj14}. 

\begin{figure}[t]
\includegraphics[scale=0.7]{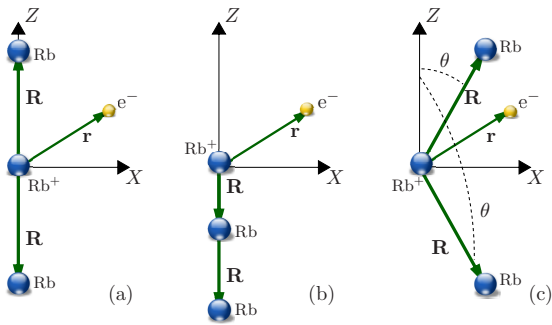}
\centering
\caption{Sketches (not to scale) of three different configurations of the triatomic molecule formed by a
 Rydberg atom and two ground-state atoms in the $XZ$-plane.} 
\label{fig:configurations}
\end{figure}

In the present work, we explore, to our knowledge for the first time, the impact of an external electric
field on the electronic structure of a triatomic ultralong-range molecule formed by a Rydberg atom and two neutral ground-state atoms.
The two ground-state atoms and the \ry core are described as point particles, 
and we restrict this study to the low-energy regime approximating the interaction between the Rydberg 
electron and each neutral atom by the  Fermi pseudopotential~\cite{fermi34,Omont}.
Compared to previous studies about the electronic structure of these triatomic \ry molecules~\cite{rost06,rost09}, 
we include both the $s$-wave and $p$-wave interactions of the Rydberg electron and the ground-state atoms. 
Our focus is on three different configurations: two collinear and one planar arrangement of the atoms.
For the collinear configurations, we consider the symmetric case where the two ground-state atoms are
located on different sides of the positively charged \ry core and at the same distance from it,
see~\autoref{fig:configurations}~(a), and an asymmetric case with the atoms being
located on the same side of the Rb$^+$ core and at different distances,  see~\autoref{fig:configurations}~(b). 
We also analyze a planar configuration where the two straight lines 
connecting the spatial positions of each ground-state atom and the \ry core form an angle smaller than $\pi$,
thereby assuming that the distances of the two ground-state atoms from the \ry core are the same, 
see~\autoref{fig:configurations}~(c).

We investigate the molecular states of collinear and planar geometries by analyzing their
adiabatic potential  curves (APC) and surfaces (APS) within the Born-Oppenheimer approximation, respectively. 
For the above-mentioned three configurations, we encounter a variety of different molecular states with deep potential
wells leading to bound ULRMs. We explore in particular the impact of an additional electric field on these adiabatic potentials
and show that the bound character of the molecular states increases in case of the linear configuration.

The paper is organized as follows. The Hamiltonian of the triatomic \ry molecule
in an external electric field is described in~\autoref{sec:hamil}. We analyze
the electronic structure of the linear and planar configurations 
in~\autoref{sec:linear_rb3} and ~\autoref{sec:planar_rb3}, respectively,
where we also explore the impact of a static electric field on the
corresponding molecular states. The conclusions are provided in~\autoref{sec:con}.

\section{The molecular  Hamiltonian}
\label{sec:hamil}
\begin{figure}[t]
\centering
\includegraphics[scale=0.8]{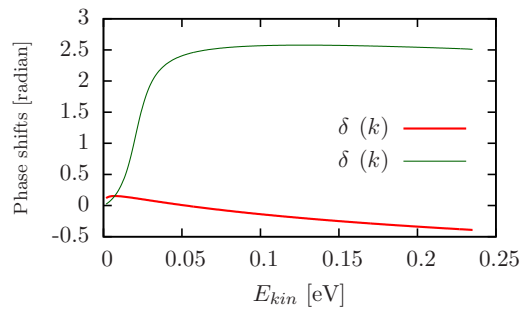}
\caption{Energy-dependent triplet phase shifts, $\delta_{0}$ (k) and $\delta_{1}(k)$, for 
the $s$- and $p$-wave scattering of an electron from the ground-state Rb atom as a function of the electron 
kinetic energy $E_{kin}$.
\label{fig:phase_shift}}
\end{figure}

We consider a triatomic  \ry  molecule formed by a \ry atom and two ground-state neutral atoms
%, \ie, two atoms, 
in a static electric field. 
It is assumed that the two ground-state atoms and the \ry core can be described as point particles
and we fix  the center of the  Laboratory Fixed Frame (LFF) at the  position of the ionic core. 
Our study focuses on the low-energy regime and  the interaction between the Rydberg electron and a 
neutral atom is described  by the  Fermi pseudopotential~\cite{fermi34,Omont}:
\begin{eqnarray}
\label{eq:fermi_ps_pot}
V(\mathbf{r},\mathbf{R}_i) = &
2\pi A_{s}[k(R_i)]\delta (\mathbf{r}-\mathbf{R}_i)\\ \nonumber
&+
6\pi A^{3}_{p}[k(R_i)] \overleftarrow{\triangledown}\delta (\mathbf{r}-\mathbf{R}_i)\overrightarrow{\triangledown}
\end{eqnarray}
where $\mathbf{r}$ and $\mathbf{R}_i=(R_i,\theta_i,\phi_i)$  are the positions of the \ry electron and neutral 
atoms with respect to the \ry  core,  respectively, with $i=1,2$.  
The energy-dependent triplet $s$- and $p$-wave  scattering lengths are given by 
$A_{s}(k)=-\tan[\delta_{0}(k)]/k$ and $A^{3}_{p}(k)=-\tan[\delta_{1}(k)]/k^{3}$, 
with  $\delta_{l}(k)$, $l=0,1$ being the corresponding phase shifts. 
The kinetic energy of the \ry electron at the collision point with the neutral atom, $R_i$, can be approximated
 by  the semiclassical expression
$E_{kin}=k^{2}/2=1/R_i-1/2n^{2}$, with $n$ being the principal quantum number of the \ry electron. The 
 energy-dependent phase shifts $\delta_{l}$ versus the  kinetic energy $E_{kin}$  of the electron 
 are presented in~\autoref{fig:phase_shift}.

The Hamiltonian of this  triatomic  \ry molecule reads 
\begin{equation}
\label{eq:full_hamiltonian}
H = T_n+H_{el}+
V(\mathbf{r},\mathbf{R}_{1})+V(\mathbf{r},\mathbf{R}_{2}),
\end{equation}
where $T_n$ is the nuclear kinetic energy operator for all relative nuclear motions.
The second term stands for the Hamiltonian of the \ry electron in an external electric field 
\begin{equation}
\label{eq:electron_hamiltonian}
 H_{el} = H_{0}+\mathbf{F}\cdot\mathbf{r} =  
 \frac{\mathbf{p}^{2}}{2m_{e}} +V_{l}(r)+\mathbf{F}\cdot\mathbf{r}
\end{equation}
with $m_{e}$ being  the electron mass, $\mathbf{p}$ its  relative momentum, and 
$V_{l}(r)$  the angular momentum $l$-dependent model potential~\cite{msd94}.
The external electric field is taken parallel to the   LFF  $Z$-axis 
  $\mathbf{F}=F\mathbf{Z}$, with $F$ being its field strength.

Within the framework of the Born-Oppenheimer approximation, 
whose validity is well justified for the ULRM given the typical energy scales of
the \ry electron as compared to the vibrational states, 
the total wave function factorizes in two  parts that describe 
 the electronic and nuclear motions of the \ry trimer. The 
total wave function can then be written as 
$\Psi(\mathbf{r},\mathbf{R}_{1},\mathbf{R}_{2})=\psi(\mathbf{r};\mathbf{R}_{1},\mathbf{R}_{2})\phi(\mathbf{R}_{1},\mathbf{R}_{2})$,
where  $\psi(\mathbf{r};\mathbf{R}_{1},\mathbf{R}_{2})$  and $\phi(\mathbf{R}_{1},\mathbf{R}_{2})$ 
are the adiabatic electronic and nuclear wave functions, respectively. 
Note that we focus within this work on the electronic structure of the triatomic  ULRM in particular
in the presence of an electric field. 
Thus, the Schr\"odinger equation for the electronic motion for fixed nuclei is  given by
\begin{eqnarray}
\left[\right.H_{0}+\mathbf{F} \cdot\mathbf{r}+&V(\mathbf{r},\mathbf{R}_{1})
+V(\mathbf{r},\mathbf{R}_{2})\left.\right]\psi_{i}(\mathbf{r};\mathbf{R}_{1},\mathbf{R}_{2})\nonumber\\
&=\epsilon_{i}(\mathbf{R}_{1},\mathbf{R}_{2})\psi_{i}(\mathbf{r};\mathbf{R}_{1},\mathbf{R}_{2})
\label{eq:sch_ry_ele}
\end{eqnarray}
with  $\epsilon_{i}(\mathbf{R}_{1},\mathbf{R}_{2})$ being the  adiabatic potential energy surface which depends
 on the position of the two  atoms $\mathbf{R}_{1}$ and $\mathbf{R}_{2}$. 
For fixed positions of the ground-state atoms,  
we solve the adiabatic electronic Schr\"odinger equation \ref{eq:sch_ry_ele} expanding
the  electronic wave function $\psi(\mathbf{r};\mathbf{R}_{1},\mathbf{R}_{2})$ 
 in the basis formed by the field-free \ry  electron wave functions
$\chi_{nlm}(\mathbf{r})=R_{nl}(r)Y_{lm}(\vartheta,\varphi)$, where 
$R_{nl}(r)$ is the radial wave function and 
$Y_{lm}(\vartheta,\varphi)$ the spherical harmonics, and $n$, $l$ and $m$ are the principal,
orbital and magnetic quantum numbers, respectively. Note that
 $H_{0}\chi_{nlm}(\mathbf{r})=E_{nl}\chi_{nlm}(\mathbf{r})$  with $E_{nl}$ being the \ry electron field-free 
eigenenergy.

We consider a triatomic Rydberg molecule formed by three  rubidium
atoms, Rb$_3$, two of them in the ground state and the third one in a \ry excited state.
Our analysis focuses on the molecular electronic states evolving from the \ry degenerate 
manifold Rb($n=35$, $l\ge3$). 
Thus, to solve the electronic Schr\"odinger equation \ref{eq:sch_ry_ele}, we include in the basis set  the 
degenerate 
manifold  Rb($n=35$, $l\ge3$) and the energetically closest neighboring Rydberg levels $38s$, $37p$ and 
$36d$. We neglect through the quantum defect of the $35f$ Rydberg state.

 %%%%%%%%%%%%%%%%%%%%%%%
\section{The linear triatomic \ry molecule}
\label{sec:linear_rb3}

\subsection{The symmetric linear configuration}

 \begin{figure}[h]
 \includegraphics[scale=0.8,angle=0]{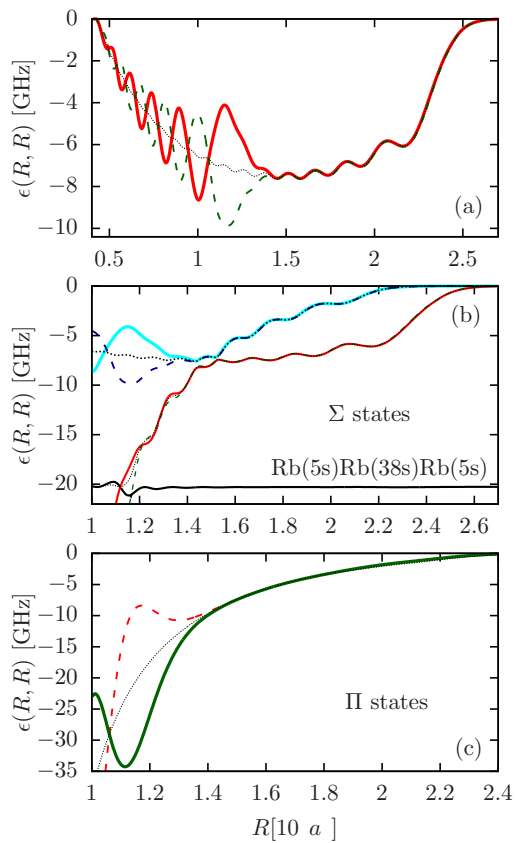}
\centering
 \caption{
 Symmetric linear triatomic ULRM with the ground-state atoms  located 
along the LFF $Z$-axis with $\theta_{1}=0$ and $\theta_{2}=\pi$, and at equal separations from the  Rb$^+$
core, \ie,   $R_{1}=R_{2}=R$. 
Adiabatic electronic potentials  as a function of the distance $R$ between the \ry core and the
neutral atoms: 
(a)  obtained including only the $s$-wave interaction, 
(b) and (c)   taking into 
account the $s$- and  $p$-wave interactions and for $\Sigma$ and $\Pi$ molecular symmetry, 
respectively.   
The gerade and ungerade symmetry curves are plotted with solid and dashed lines, respectively. 
In panel (b),  the APC evolving from the Rb($38s$) \ry state is also shown. 
For the sake of completeness, we plot the APC of  the diatomic  ULRM (dotted lines)  obtained including 
(a)   only the $s$-wave interaction, 
(b) and (c)   both the $s$- and  $p$-wave interactions and for $\Sigma$ and $\Pi$ molecular symmetry, 
respectively.   
The zero energy has been set to the energy of the field-free  Rb($n =35, l\ge3$) degenerate manifold.}
   \label{fig:FF_linear_1_35_l}
 \end{figure} 
 
We start by considering the symmetric linear triatomic ULRM  presented~in \autoref{fig:configurations}~(a), 
with the neutral atoms located
along the LFF $Z$-axis 
at different sides  of the \ry core and at the same distance, \ie, $R_1=R_2=R$ and $\theta_1=0$,
$\theta_2=\pi$. 
This \ry molecule has been previously investigated in field-free space by 
modeling the interaction between the \ry electron and the neutral atoms via  the 
$s$-wave scattering pseudopotential and neglecting the contribution of the 
$p$-wave scattering~\cite{rost06,rost09}. 
For this triatomic system, the  APCs with $\Sigma$ molecular symmetry 
(and $s$-wave interaction only) 
are presented in~\autoref{fig:FF_linear_1_35_l}~(a),  where the 
 APC of a diatomic ULRM with the ground state atom 
located on the LFF $Z$-axis is also shown. 
For the triatomic ULRM, we observe that   two adiabatic potential curves with gerade and 
ungerade symmetry split away from the Rb($n = 35$, $l\ge 3$) degenerate manifold.
In Ref.~\cite{rost06}, it is shown that the gerade (ungerade) APC can be written as a sum in terms 
of the probability densities of \ry electronic states with even (odd) angular momentum $l\ge 3$. 
These two APC oscillate around the adiabatic electronic curve of the diatomic ULRM, and, in particular,
exhibit deeper potential wells, which indicates a higher degree of stability. 
At large separations between the atoms and Rb$^+$, these potential curves converge to the 
APC of a diatomic  Rydberg molecule with only 
one  ground-state atom, see~\autoref{fig:FF_linear_1_35_l}~(a),
because the sums of even-$l$ and odd-$l$ probability densities that  contribute to the gerade and ungerade
molecular  states, respectively,  are very similar as $R$ increases~\cite{rost06}. 
Based on previous studies which show that the $p$-wave scattering pseudopotential plays a crucial role on the 
electronic structure of diatomic ULRMs~\cite{hamilton02,kurz13,fabrikant02}, we present here the
electronic structure of this triatomic ULRM 
 including both the $s$ and $p$-wave interactions of the \ry electron and the ground-state atoms.
The molecular electronic potentials  with $\Sigma$ and $\Pi$ symmetry of the triatomic ULRM are shown
in Figures.~\ref{fig:FF_linear_1_35_l}~(b) and (c), respectively. For the sake of comparison, the APCs of a diatomic ULRM with the ground state atom located on the LFF $Z$-axis are also presented.

Let us first analyze the $\Sigma$ molecular levels. 
For the triatomic ULRM, the $p$-wave interaction provokes that 
two additional  potentials with $\Sigma$  molecular symmetry are shifted away from  the  Rb($n = 35$, $l\ge 3$) 
degenerate manifold. 
The resonance of the  $p$-wave scattering  length at $R\approx780~a_0$ significantly affects  the APC 
and their slope becomes pronounced for $R\lesssim 1200~a_0$. 
Indeed,  two of these APCs suffer avoided crossings with 
 the adiabatic state evolving from the non-degenerate \ry level  Rb($38s$), whose energy,
 on the scale of the figure, remains 
 approximately constant  for larger values of $R$.  
The  $p$-wave and $s$-wave dominated $\Sigma$-APCs suffer several avoided crossings close to 
the internuclear distance $R\approx 1500~a_0$.
The oscillatory behavior of these adiabatic potentials is due to the highly oscillatory character of the \ry 
electron 
wave function in the Rb($n=35$, $l\geqslant 3$) state. 
The ground-state atoms, considered to a good approximation as 
point particles,  probe   locally in space the  electronic wave function of the highly excited \ry atom. 
At large separations between the \ry core and the ground-state atoms, the  $s$-wave 
($p$-wave) 
dominated $\Sigma$ molecular states become degenerate and converge to the $s$-wave ($p$-wave)   
$\Sigma$-APC of the  diatomic  ULRM.
For these large values of $R$, the electronic structure  of the triatomic ULRM  is composed of 
the molecular states of two diatomic ULRMs that share the same \ry core and  one has the ground-state 
atom at $\theta=0$ whereas  the other one has it at $\theta=\pi$. 

We focus now on the  APCs with $\Pi$ molecular symmetry.
The  $p$-wave interaction is responsible for the adiabatic potentials with $\Pi$ molecular 
symmetry, cf.~\autoref{fig:FF_linear_1_35_l}~(c). In contrast to 
the  $\Pi$ molecular levels of  the diatomic ULRM, these two APCs show potential wells that 
could accommodate  several vibrational bound states. 
For large internuclear distances, these two adiabatic electronic
 potentials become degenerate and equal to the corresponding potentials of a diatomic ULRM.

 \begin{figure}[t]
 \includegraphics[scale=0.8,angle=0]{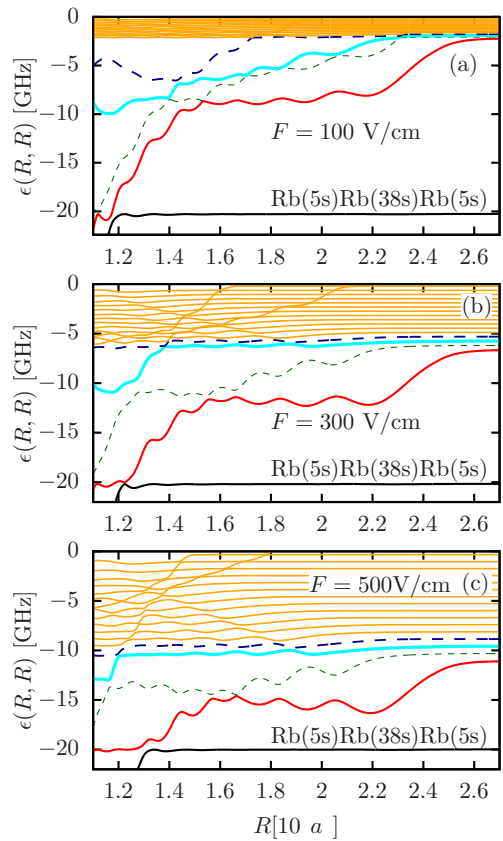}
\centering
 \caption{For the symmetric linear ULRM:  
molecular states  with $\Sigma$ symmetry evolving from the degenerate manifold Rb($n=35$, $l\ge 3$)  versus the interatomic distance $R$  of the ground-state atoms from 
 the Rydberg core  in an external electric field of strength 
 (a)  $F=100$~V/m,  (b) $F=300$~V/m,  and (c) $F=500$~V/m.
The APCs have been derived including both the $s$-wave and $p$-wave interaction.  
  The zero energy has been set to the energy of the field-free  Rb($n =35$, $l\ge3$) degenerate manifold.}
\label{fig:linear_trimer_field}
 \end{figure}
Due to their large dipole moments, \ry atoms are extremely sensitive to weak static electric fields, and, as 
consequence, the level structure of the 
ULRM is also significantly affected. In~\autoref{fig:linear_trimer_field}~(a), (b) and (c), we present 
the APCs with $\Sigma$ molecular symmetry in a static electric field
of strength $F=100$~V/m, $300$~V/m, and $500$~V/m, respectively. 
Due to the interaction with the dc field,
additional molecular states are shifted from  the field-free Rb($n = 35$, $l\ge 3$) \ry manifold,
which reflects that the Stark interaction couples states with field-free 
orbital numbers $l$ and $l\pm1$. 
The 	electric field couples the adiabatic electronic states with gerade and ungerade symmetry
and the degeneracy at large internuclear distances of these APCs is lifted. 
For  $F=100$~V/m, the overall energy of one of the  states
 with $s$-wave ($p$-wave) dominated character decreases, whereas the energy of the second
one  increases approaching the adiabatic electronic states from the Rb($n = 35$, $l\ge 3$) manifold.
By further increasing  $F$, the energies of all APCs show a decreasing behaviour.
For  $F=500$~V/m, two of the APCs  are merged with the field-dressed adiabatic electronic
potentials that are split from the field-free Rb($n = 35$, $l\ge 3$) \ry manifold, and we encounter several avoided 
crossings, see~\autoref{fig:linear_trimer_field}~(c). 
Thus, only two molecular states remain energetically well separated from the field-dressed levels
evolving from the  Rb($n = 35$, $l\ge 3$) \ry manifold. 
The avoided crossing  between the lowest-lying  APCs from the \ry trimer
Rb($5s$)Rb($n = 35$, $l\ge 3$)Rb($5s$) and the molecular state from Rb($5s$)Rb($38s$)Rb($5s$) becomes 
broader as $F$ is increased. 
Note that  due to the quadratic Stark shift of the $38s$ \ry state, the APC of Rb($5s$)Rb($38s$)Rb($5s$)
is very weakly affected by the electric field. 
Due to the interplay between the interaction of the \ry atom with the two neutral atoms and with the 
external electric field, the change in energy of the APCs depends on the internuclear separation 
between  the \ry core and  the two ground-state atoms. 
At large internuclear distances, when the interaction with the two ground-state atoms becomes 
negligible,  all the APCs from the Rb$(n=35,l\ge3$) \ry manifold are shifted linearly in energy  with the 
dc field strength, which corresponds with the Stark shift of the $(n=35,l\ge3$)  \ry manifold of an isolated Rb atom.

\begin{figure}[t]
  \centering
\includegraphics[scale=0.8,angle=0]{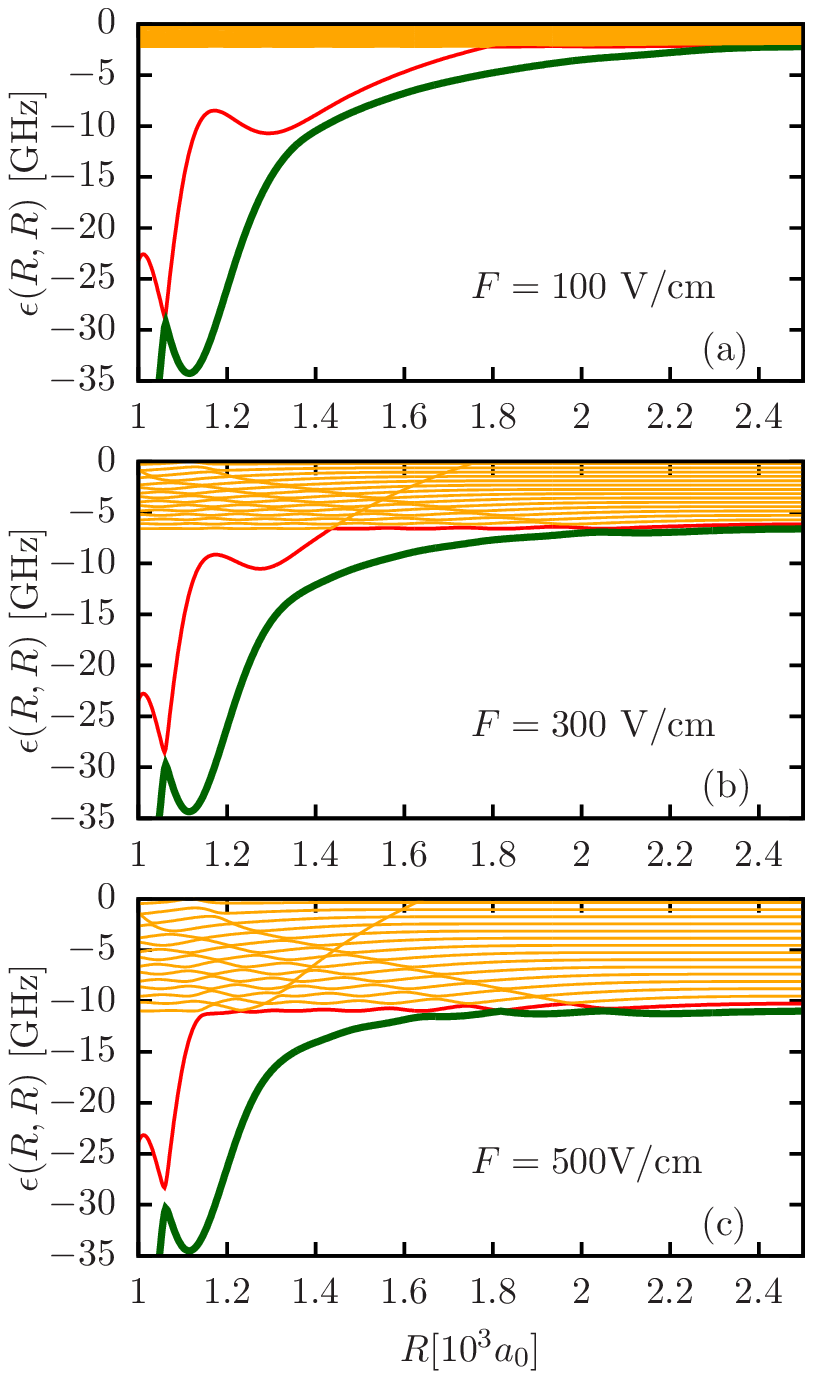}
 \caption{For the symmetric linear ULRM: electronic states with $\Pi$ molecular symmetry evolving from the degenerate manifold Rb($n=35, l \ge 3$)  versus the interatomic distance $R$ 
 of the ground-state atoms from the \ry core  in an electric field of strength 
 (a)  $F=100$~V/m,  (b) $F=300$~V/m,  and (c) $F=500$~V/m.
 The APCs have been derived including both the $s$-wave and $p$-wave interaction.  
  The zero energy has been set to the  field-free   energy of the Rb($n =35$, $l\ge3$) degenerate manifold.}
\label{fig:linear_pi_trimer_field}
 \end{figure}

Analogous results are found for  the field-dressed molecular states with $\Pi$ molecular symmetry, 
see~\autoref{fig:linear_pi_trimer_field}.  For large internuclear separations $R\gtrsim1400~a_0$, 
the degeneracy  of the two $\Pi$ 
adiabatic  electronic states is lifted.
For $F=100$~V/m, one of the APCs increases in  energy  compared to its field-free value,
whereas the other one decreases,~see~\autoref{fig:linear_pi_trimer_field}~(a), whereas
for $F=300$~V/m and $F=500$~V/m the energies of all APCs are reduced.  
In addition, the crossing of the field-free APCs at  $R\approx 1060~ a_0$ becomes an avoided crossing 
in the presence of the dc field. 
These $\Pi$ molecular states are weakly  affected by the external field for 
$1000~a_0\lesssim R\lesssim 1250~a_0$, 
their Stark shifts depend quadratically on the  field strength and  are hardly visible on the 
scale of~\autoref{fig:linear_pi_trimer_field}. 
For instance, the energy of the minimum appearing at $R\approx 1115~a_0$ for
the lowest lying $\Pi$-APC  
is shifted  $0.3$~GHz from  its field-free value for  $F=500$~V/m.
As $F$ is increased, the field-free highest-lying APC gets mixed with the additional field-dressed APCs 
evolving from the  Rb$(n=35,l\ge3)$ manifold suffering several avoided crossings with them. 
Again, at large internuclear distances, the Stark  shifts of 
all the APCs from the Rb$(n=35,l\ge3$) \ry manifold depend linearly  on $F$.
 
 \subsection{The non-symmetric linear configuration}

\begin{figure}[t]
 \centering
\includegraphics[scale=0.8,angle=0]{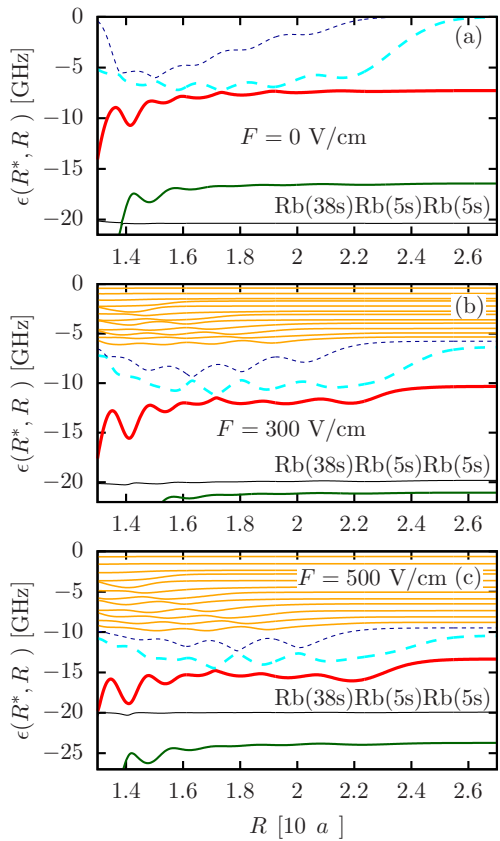}
 \caption{Asymmetric linear  ULRM with the two ground-state atoms located at the same
 side of the \ry core, \ie, $\theta_{1}=\theta_{2}=\pi$
 and one fixed at $R_1^*=1200~a_0$. 
 APCs with $\Sigma$ molecular symmetry evolving from the degenerate manifold 
 Rb($n=35$, $l\ge 3$)  
 versus the separation $R_2$ between the second atom and Rb$^+$, 
 for an electric field of strength 
 (a) $F=0$~V/m, (b) $F=300$~V/m,  and (c) $F=500$~V/m.
 The APCs have been derived including both the $s$-wave and $p$-wave interaction.  
The zero energy has been set to the  field-free    energy of the Rb($n =35$, $l\ge3$) degenerate manifold.}
\label{fig:FF_linear_same_side}
 \end{figure}

\begin{figure}[t]
\centering
\includegraphics[scale=0.8,angle=0]{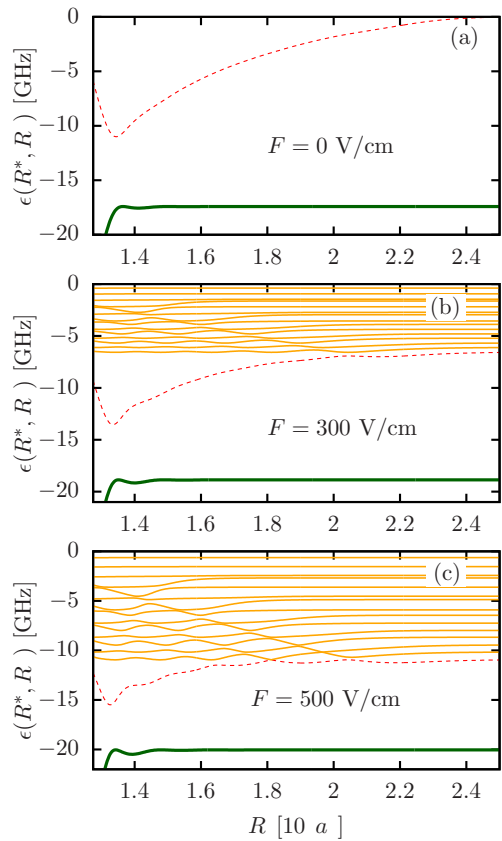}
 \caption{Asymmetric linear  ULRM,
 with the two ground-state atoms located at the same
 side of the \ry core, \ie, $\theta_{1}=\theta_{2}=\pi$
 and one fixed at $R_1^*=1200~a_0$. 
  APCs  with $\Pi$ molecular symmetry evolving 
 from the degenerate manifold Rb($n=35$, $l\ge 3$) as a function of 
distance $R_2$ between the \ry core and the second atom 
 for an electric field of strengths 
 (a) $F=0$~V/m, (b) $F=300$~V/m,  and (c) $F=500$~V/m.
 The APCs have been derived including both the $s$-wave and $p$-wave interaction.  
The zero energy has been set to the  field-free   energy of the Rb($n =35$, $l\ge3$) degenerate manifold.}
 \label{fig:FF_linear_same_side_pi}
 \end{figure}

We consider now a linear triatomic  \ry molecule with the two ground-state atoms located along the LFF 
$Z$-axis and  at the same side of the \ry core,  \ie, $\theta_{1}=\theta_{2}=\pi$. 
A sketch of this configuration is presented in~\autoref{fig:configurations}~(b).
The position of one of the atoms  is fixed at the $R_1=1200~a_0$, 
whereas the distance of the second one from the Rb$^+$ core, $R_2$, varies for 
$R_2>R_1$.  This condition allows us to neglect the interaction between both ground-state atoms
 because their vibrational wave functions don't overlap in space.
Note that the distance of the first atom from the Rb$^+$ core has been arbitrarily fixed to $R_1=1200~a_0$,  
and that qualitatively similar results are obtained for other values of $R_1$. 
The  $\Sigma$ and $\Pi$ molecular states 
are plotted versus the internuclear separation  $R_2$ in~\autoref{fig:FF_linear_same_side}
and~\autoref{fig:FF_linear_same_side_pi}, respectively.
For the electronic structure with $\Sigma$  molecular symmetry, 
the presence of the second atom provokes that two extra adiabatic electronic states  are split away from the  
Rb($n=35$ and $l\ge 3$)  degenerate manifold. 
These two APCs show an oscillatory behaviour superimposed to a broad potential well over
the regions $1400~a_0\lesssim R_2\lesssim  2350~a_0$ and 
$1300~a_0\lesssim  R_2\lesssim 2700~a_0$ for the $p$-wave 
and $s$-wave dominated states, respectively. 
These APCs can accommodate many vibrational levels, 
with spacing $\sim 150$MHz approximately, some of the them having   
spatial  extensions of a few hundreds Bohr radii.  
Several avoided crossings are encountered between these two adiabatic electronic states, and also between 
the $s$-wave dominated state and the $p$-wave dominated state evolving from the \ry diatomic molecule.
For large values of   $R_2$, the interaction with the second atom becomes negligible and 
the triatomic \ry molecule becomes an effective  diatomic system. 
In this case, the two energetically lowest $\Sigma$-symmetry  APCs have  the energies of the 
corresponding  electronic states of the  diatomic \ry molecule
with the ground-state atom located at $R_1=1200~a_0$; 
whereas the other two APCs approach zero energy.
Due to the second atom,  an extra  $\Pi$ molecular state is shifted from the 
Rb($n=35$ and $l\ge 3$)
\ry manifold, see~\autoref{fig:FF_linear_same_side_pi}~(a), which shows a deep well with a minimum at $R\approx 1345~a_0$.
This APC  is deep enough to  accommodate several vibrational bound states, where the triatomic molecule 
would exist. 
The lowest-lying molecular state of this symmetry has a shallow potential well  of approximately $150$~MHz deep at $R_2\approx1410~a_0$.
For $R\gtrsim1500~a_0$, the lowest lying $\Pi$-APC shows a constant behaviour
with the energy of the corresponding APC of the diatomic ULRM.

In the presence of an external  electric field, the energy of the APCs with $\Sigma$ and $\Pi$ molecular 
symmetries  decreases as $F$ is increased, see~\autoref{fig:FF_linear_same_side}
and ~\autoref{fig:FF_linear_same_side_pi}. 
We start analyzing the results for the electronic states with $\Sigma$ molecular symmetry.
The APC of the  Rb($38s$)Rb($5s$)Rb($5s$) trimer  is weakly affected by the electric field due to the quadratic 
Stark effect of the  Rb($38s$) \ry state.
The lowest-lying APC evolving  from the Rb($n=35, l\ge 3$) manifold has an energy
smaller than the Rb$(38s)$Rb($5s$)Rb($5s$) electronic state for $F=300$~V/cm and $500$~V/cm. 
The second-lowest-lying APC evolving  from the Rb($n=35, l\ge 3$) manifold also decreases in energy and
its outermost well becomes more pronounced and shows an increasing depth as $F$ is increased.
For the two extra adiabatic electronic states with  $\Sigma$ symmetry, which appear  due to the second 
atom, the electric field provokes that their broad wells become less deep whereas the 
superimposed oscillatory behaviour remains,  see Figures~\ref{fig:FF_linear_same_side}~(b) and (c). 
The Stark effect breaks the  degeneracy of the field-free degenerate adiabatic electronic states 
and extra APCs are shifted from the field-free degenerate manifold Rb($n=35, l\ge3$). 
These extra molecular levels show an oscillatory behavior for $R_2\lesssim 2000~a_0$, which provokes many 
avoided crossings among  them,  and  for larger values of  $R_2$, they possess a constant energy
approximately.

Similar results are observed for the adiabatic energies with $\Pi$ molecular symmetry, 
see~\autoref{fig:FF_linear_same_side_pi}. 
The lowest-lying molecular state 
evolving  from the Rb($n=35, l\ge 3$) \ry manifold suffers a quadratic Stark effect
and, therefore, is weakly  affected by the weak electric field. The second lowest-lying potential decreases in 
energy as $F$  increases and the depth of the pronounced wells is significantly reduced,
but remains  a few GHz deep and can accommodate vibrational bound states. An increasing number of
avoided crossings for large values of $R_2$ with increasing field strength is encountered. 

\section{The planar  triatomic \ry molecule}
\label{sec:planar_rb3}

\begin{figure*}[t]
 \centering
\includegraphics[scale=0.8,angle=0]{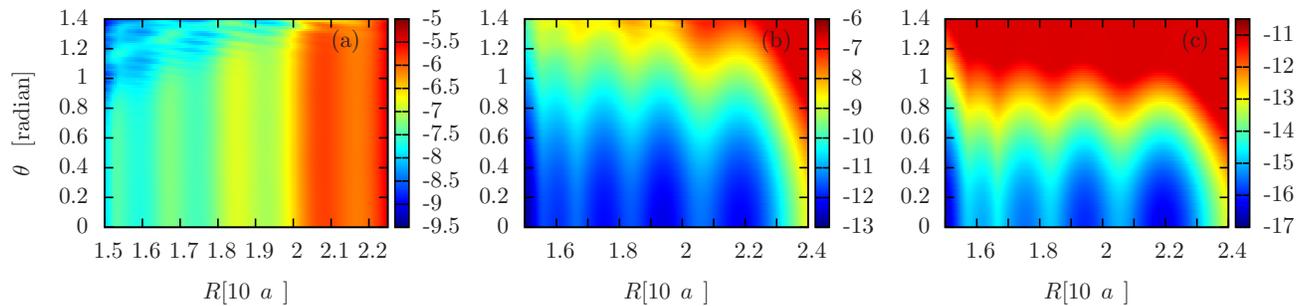}
 \caption{For the planar ULRM: The lowest lying APS evolving from the  Rb($n =35$, $l\ge3$) \ry manifold
 in an external electric field parallel to the LFF $Z$ axis and field strength 
 (a) $F=0$~V/m, (b) $F=300$~V/m,  and (c) $F=500$~V/m. 
 The APSs have been derived including both the $s$-wave and $p$-wave interaction.  
 The zero energy has been set to the  field-free   energy of the Rb($n =35$, $l\ge3$) degenerate manifold.}
 \label{fig:FF_planar_side}
 \end{figure*}

In this section, we  consider a planar ULRM with the neutral atoms located in
 the LFF $XZ$-plane, \ie, $\phi_1=\phi_2=0$,
and $\theta_2=\pi-\theta_1$ with 
$0\le \theta_1<\pi/2$, see~\autoref{fig:configurations}~(c). For the sake of simplicity, we restrict this study to 
the configuration with the ground-state atoms located at the same distance from the \ry core, \ie, 
$R_1=R_2=R$. Note that we impose the conditions $0\le\theta_1<\pi/2$  and  $\pi/2<\theta_2\le \pi$  
to avoid the spatial overlap of the vibrational states of the two ground-state atoms, such 
 that  the interaction between them can  be neglected.
The two energetically lowest lying APS evolving from the Rb$(n=35,l\ge3$) \ry manifold are
shown as a function of  $R$ and  $\theta_1$ in~\autoref{fig:FF_planar_side}. 

The field-free APS shows a weak dependence on the polar angle $\theta_1$ for different values of 
$R$, cf.~\autoref{fig:FF_planar_side}~(a).
For $R\gtrsim 2000~a_0$,  these field-free APSs are almost independent of the angle $\theta_1$,
whereas for smaller internuclear separations, we encounter 
a smooth dependence on $\theta_1$ for  $\theta_1\gtrsim 0.8$. This is due to the spatial proximity of the 
two ground-state atoms.
By adding an electric field, the APS strongly depends on the polar angle $\theta_1$. 
The electric field parallel to the $Z$ axis  favours the linear configuration of the ULRM  
with  both ground-state atoms located along the $Z$ axis, see Figures~\ref{fig:FF_planar_side}~(b) and (c).
For this APS, the deepest well is shifted towards larger values of $R$ as $F$
is increased, and it is located  at $R\approx 2200~a_0$, $\theta_1=0$ and 
$\theta_2=\pi$  for $F=500$~V/m. 
This molecular curve approaches to a constant energy, given by the corresponding \ry state of the 
field-dressed Rb for large internuclear separations and large values of $\theta_1$.
For $F=500$~V/m,  this effect is observed for 
$\theta_1\gtrsim1.1$, and  its limit is the energy the lowest-lying  field-dressed  state 
of the \ry manifold $(n=35,l\ge3$) of an isolated Rb atom in an electric field.

%%%%%%%%%%%%%%%%%%%%%%%%%%%%%%%%%%%%%%
\section{Conclusions}
\label{sec:con}

We have investigated ultralong-range triatomic \ry molecules formed by a \ry rubidium atom and two
ground-state Rb atoms in the  presence of an external electric field.  
This is, to our knowledge, the first investigation of such a triatomic ULRM in an electric field. 
The symmetric and  asymmetric linear configurations, as well as 
a planar one with the neutral atoms located at the same distance from the \ry core have been explored. 
We have performed an analysis of the electronic structure for the Rb \ry atom in the 
 degenerate manifold ($n=35, l\ge 3)$.
 For the  linear configurations,  in the absence of the electric field several molecular  states, 
 with $\Sigma$ and $\Pi$ symmetry, 
 are split from the Rb($n=35, l\ge 3)$ manifold,  showing an oscillatory behaviour with many wells  
 accommodating vibrational levels where the \ry molecule would exist.
 For the planar configuration, we encounter that the molecular states only show a significant dependence on
 $\theta_1$ and $\theta_2$ as the two ground-state atoms approach each other.
The electric field favours the symmetric linear  configuration
and strengthen the bound state character of molecular states, and vibrational states
 with the spatial extensions of  a few hundreds Bohr radii could appear.

An inmediate extension of this work would be to investigate the electronic structure and 
vibrational properties of an ultralong-range  triatomic molecule formed from 
a Rb \ry atom in a non-degenerate   state with orbital quantum number $l\le 3$. 
For diatomic  \ry  molecules, it has been shown that the 
Rydberg-atom fine structure and the hyperfine coupling of the ground-state atom
significantly alter the adiabatic electronic potentials evolving from the \ry atom in such excited 
states~\cite{Sassmannshausen15,Anderson14}.
By including these couplings for a triatomic \ry molecule, the potential energy surfaces depend on the 
 hyperfine states of the ground-state atoms as well as on the fine structure of the \ry atom.
 As a consequence, the complexity of the molecular states would be significantly enhanced.

\begin{acknowledgments}
R.G.F.  and J.A.F.  acknowledge financial support by the Spanish project FIS2014-54497-P (MINECO), 
and by the Andalusian research group FQM-207.
We also acknowledge financial support by the Initial Training Network COHERENCE of the European 
Union FP7 framework. We would like to thank Markus Kurz  and Christian Fey for fruitful discussions. 
\end{acknowledgments}

\end{document}